# hBN alignment orientation controls moiré strength in rhombohedral graphene

Matan Uzan[1†], Weifeng Zhi[1†], Matan Bocarsly[1†], Junkai Dong[2], Surajit Dutta[1], Nadav Auerbach[1], Niladri Sekhar Kander[1], Mikhail Labendik[1], Yuri Myasoedov[1], Martin E. Huber[3], Kenji Watanabe[4], Takashi Taniguchi[5], Daniel E. Parker[6#], and Eli Zeldov[1*]

Rhombohedral multilayer graphene hosts a rich landscape of correlated symmetry-broken phases, driven by strong interactions from its flat band edges. Aligning to hexagonal boron nitride (hBN) creates a moiré pattern, leading to recent observations of exotic ground states such as integer and fractional quantum anomalous Hall effects. Here, we show that the moiré effects and resulting correlated phase diagrams are critically influenced by a previously underestimated structural choice: the hBN alignment orientation. This binary parameter distinguishes between configurations where the rhombohedral graphene and hBN lattices are aligned near 0° or 180°, a distinction that arises only because both materials break inversion symmetry. Although the two orientations produce the same moiré wavelength, we find their distinct local stacking configurations result in markedly different moiré potential strengths. Using low-temperature transport and scanning SQUID-on-tip magnetometry, we compare nearly identical devices that differ only in alignment orientation and observe sharply contrasting sequences of symmetry-broken states. Theoretical analysis reveals a simple mechanism based on lattice relaxation and the atomic-scale electronic structure of rhombohedral graphene, supported by detailed modeling. These findings establish hBN alignment orientation as a key control parameter in moiré-engineered graphene systems and provide a framework for interpreting both prior and future experiments.

___________________________

[1]Department of Condensed Matter Physics, Weizmann Institute of Science, Rehovot 7610001, Israel
[2]Department of Physics, Harvard University, Cambridge, MA 02138, USA
[3]Departments of Physics and Electrical Engineering, University of Colorado Denver; Denver, Colorado 80217, USA
[4]Research Center for Electronic and Optical Materials, National Institute for Materials Science; 1-1 Namiki, Tsukuba 305-0044, Japan
[5]Research Center for Materials Nanoarchitectonics, National Institute for Materials Science; 1-1 Namiki, Tsukuba 305-0044, Japan
[6]Department of Physics, University of California at San Diego, La Jolla, California 92093, USA
[†]These authors contributed equally to this work
[#]danielericparker@ucsd.edu
[*]eli.zeldov@weizmann.ac.il

Van der Waals heterostructures combining graphene and hexagonal boron nitride (hBN) have emerged as a powerful platform to engineer strongly correlated and topological phases. Among these, rhombohedral-stacked multilayer graphene (RMG) stands out due to its flat, low-energy bands, which place the system in an interaction-dominated regime where strongly correlated phases appear [1,2]. As carrier density is tuned, RMG exhibits a robust sequence of metallic states with decreasing degeneracy: from a fourfold-degenerate metal at high density, to a spin-polarized twofold metal, and eventually to a fully spin- and valley-polarized onefold metal near charge neutrality [3,4]. Although the precise phase boundaries depend on external parameters such as displacement field and number of layers, this symmetry-breaking hierarchy appears remarkably robust across a broad range of RMG devices [3–7]. When RMG is aligned to hBN, the resulting moiré superlattice [8–15] introduces periodic modulation and additional symmetry breaking, leading to band reconstruction. Recent experiments have observed moiré minibands with exotic topological states, such as integer and fractional quantum anomalous Hall insulators [16–21], whose origin is under intense theoretical debate [22–27].

The moiré between hBN and RMG is typically characterized experimentally by its moiré wavelength—but this does not uniquely specify the moiré pattern. This is because both hBN and RMG break in-plane $\pi$-rotation symmetry, $C_{2z}$. As a result, rotating the hBN by 180° relative to RMG leaves the moiré wavelength unchanged, but swaps the local stacking of boron and nitrogen with respect to graphene's A and B sublattices. This defines a discrete binary degree of freedom that we refer to as *alignment orientation*. The two alignment orientations, labeled $\xi = 0$ and $\xi = 1$, are not symmetry related and therefore specify two classes of distinct physical systems. Despite a long history of careful studies on hBN alignment [9,10,13,28], previously chosen continuum model parameters critically underestimate its effect. A few studies have begun to address this point, including experimental evidence of $\xi$-dependent non-local transport [29] and crystal fields [30] in bilayer graphene. Until now, a systematic investigation of alignment orientation in correlated graphene based moiré systems has been lacking.

In this work, we show that the alignment orientation $\xi$ exerts a profound influence on correlated ground states in rhombohedral pentalayer graphene (R5LG) aligned to hBN. This difference is most apparent in the moiré-proximate regime, where an external displacement field is used to push the charge carriers towards the bottom graphene layer, proximate to the aligned hBN. Tunneling between graphene and hBN creates a moiré potential that is dominated by a particular carbon-boron tunnelling process in part of the unit cell [12]. We find that this process is essentially switched on or off by the alignment orientation, effectively creating a strong moiré potential for $\xi = 1$, but a weak moiré for $\xi = 0$. This microscopic difference ultimately leads to qualitatively distinct correlated states and phase diagrams for the two alignment orientations. We demonstrate that these differences not only distinguish between $\xi = 1$ and $\xi = 0$ devices, but also account for the main variations and inconsistencies among previously reported phase diagrams in RMG/hBN aligned devices.

To isolate the role of $\xi$, we fabricated nearly identically aligned R5LG/hBN devices that differ only in their hBN alignment orientation. Transport measurements reveal markedly different symmetry-broken phases in the electron-doped regime. Complementary SQUID-on-tip magnetometry further uncovers sharp magnetic transitions, spin and valley-polarized states, and an intervening weakly-magnetic phase that appears only in one $\xi$ configuration. These experimental results are supported by single-particle and mean-field theoretical models, which show how $\xi$ controls the single-particle band structure and consequently the hierarchy of spontaneous symmetry breaking. Our findings identify alignment orientation as a critical tuning knob in moiré-engineered graphene systems that must be accounted for to interpret past experiments and design future devices.

**Transport measurements**

Four dual-gated R5LG/hBN devices were fabricated using the dry-transfer method (Extended Data Fig. 1, Methods). In all four devices, the R5LG is nearly perfectly aligned to the bottom hBN substrate, with an estimated twist angle $\theta \approx 0.3°$. The dc voltages $V_{tg}^{dc}$ and $V_{bg}^{dc}$, applied to the top and bottom gates, enable simultaneous tuning of the carrier density $n$ and the out-of-plane displacement field $D$ (Fig. 1c). Transport measurements of longitudinal resistance $R_{xx}$ and Hall carrier density $n_H$, were performed at a temperature $T \approx 500$ mK as a function of the filling factor $\nu = 4n/n_s$ and $D$ (where $n_s$ denotes the density corresponding to four electrons per moiré unit cell). Despite their nearly identical twist angles, the devices exhibit qualitatively different behavior and naturally separate into two distinct classes (Extended Data Fig. 2). Anticipating their microscopic origin, we label these classes by their alignment orientation, $\xi = 0$ or $\xi = 1$.

Figures 1a,b show $R_{xx}(\nu, D)$ for representative $\xi = 1$ and $\xi = 0$ devices, respectively. A positive (negative) $D$ pushes electrons (holes) toward the graphene layer adjacent to the aligned hBN (moiré-proximate side, Fig. 1c). In the moiré-distant quadrants (see labels in Fig. 1a), the phase diagrams of both devices are broadly similar to unaligned rhombohedral graphene samples (Methods). In contrast, the moiré-proximate regime reveals striking differences between the two devices. These differences are especially pronounced at positive integer moiré fillings (Fig. 1e). The $\xi = 1$ device (blue curves) exhibits strong insulating states at all integer fillings $\nu = 1, 2, 3$, and $4$, with peak resistances exceeding ~100 kΩ that persist across the entire experimentally accessible range of high $D$. The insulating state at $\nu = 4$ corresponds to a single-particle band insulator with four electrons per moiré unit cell, while the states $\nu = 1, 2, 3$ reflect interaction-driven, symmetry-broken insulators common in strongly correlated moiré systems [31]. By contrast, the $\xi = 0$ device exhibits much weaker insulating features that terminate at moderate $D$. The $R_{xx}$ peak at $\nu = 1$ is substantially reduced, and those at $\nu = 2, 3, 4$ reach only a few kΩ, persisting over narrow regions of $D$. The contrast is especially stark at $\nu = 3$ and 4: for $\nu = 3$, the $\xi = 0$ device displays only a faint, sharply confined resistance peak, while at $\nu = 4$, the insulating feature, though broader in $D$, is markedly weaker than in the $\xi = 1$ device and terminates abruptly.

To further investigate the symmetry-broken phases, we measure the Hall resistance while sweeping the out-of-plane magnetic field $B_a$. Figure 1d presents representative $R_{xy}(B_a)$ traces taken at a fixed point in the $(\nu, D)$ plane marked by the red dots in Figs. 1a,b. The $\xi = 0$ device shows a pronounced anomalous Hall effect with a sharp, hysteretic jump in $R_{xy}$ across $B_a = 0$ T, signaling spontaneous time-reversal symmetry breaking and the emergence of orbital magnetization. In contrast, the $\xi = 1$ device at the same $(\nu, D)$ shows a linear, non-hysteretic $R_{xy}(B_a)$ with zero value at $B_a = 0$ T, consistent with a nonmagnetic ground state.

These contrasting behaviors reveal a fundamental difference in the spontaneous symmetry breaking between the two devices, despite their nearly identical twist angles. The discrepancy is most pronounced on the moiré-proximate side, where the influence of the moiré potential is strongest— pointing to underlying moiré structure as the likely origin. Given the near-identical twist angles across all devices, we attribute the observed differences between the two classes of devices to the alignment orientation, which determines the effective strength of the moiré potential, as discussed next.

**Single-particle theory**

To investigate the impact of alignment orientation, we undertake a theoretical analysis of the single-particle band structure of R5LG aligned with hBN. We model the system using a continuum approach that incorporates the moiré superlattice potential, interlayer coupling, sublattice asymmetry, and phenomenological lattice relaxation [8,9,11,12,32]. This framework enables us to evaluate how the alignment orientation affects the resulting miniband structure.

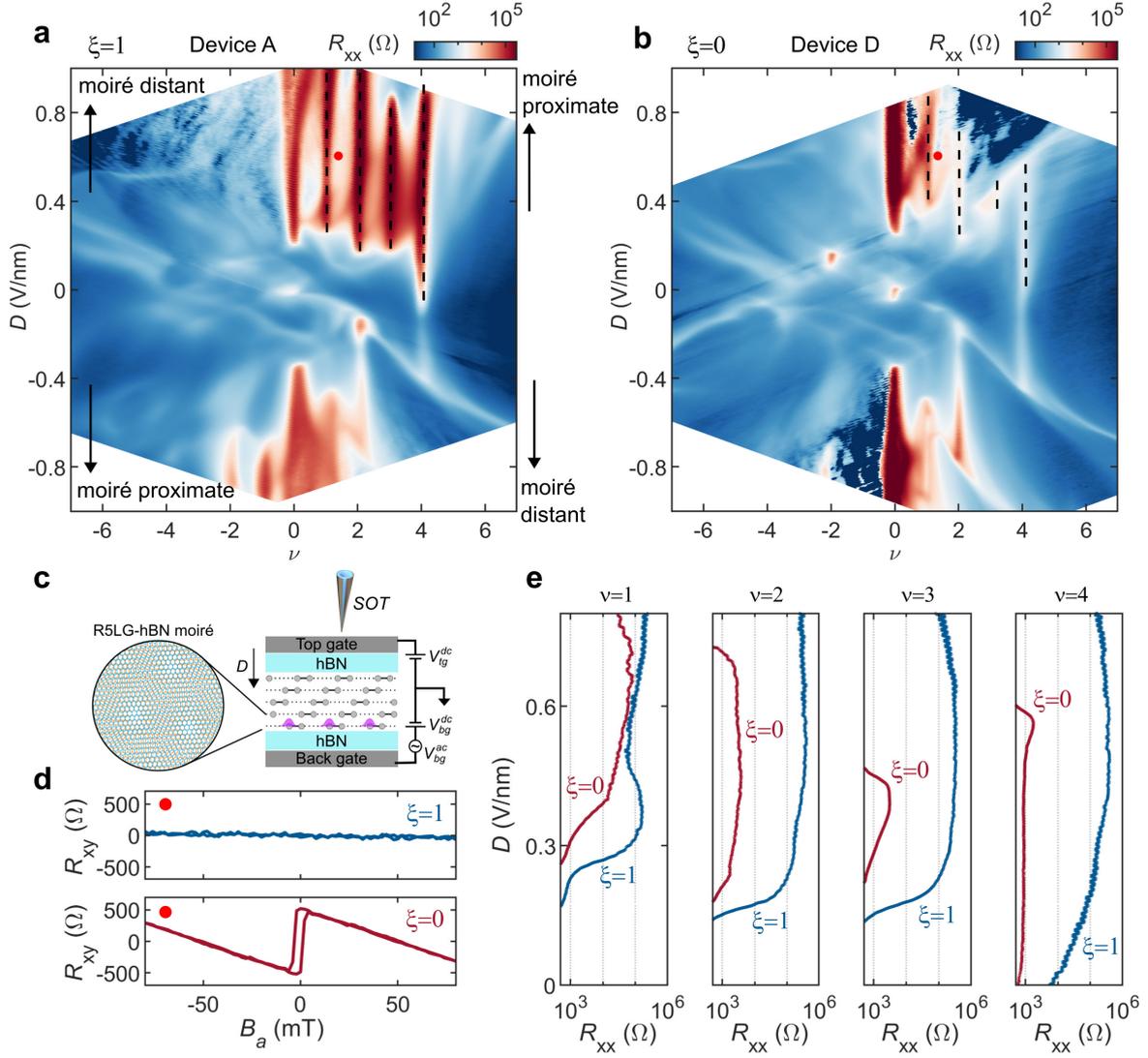

**Fig. 1. Transport measurements of R5LG/hBN devices. a**, $R_{xx}$ versus $\nu$ and $D$ measured at $B_a = 0$ T and $T = 500$ mK in $\xi = 1$ device. Moiré-proximate (distant) quadrants, where the active carriers are pushed towards (away from) the aligned hBN, are labeled. In the electron-doped, moiré-proximate regime (upper right quadrant), resistance peaks appear at charge neutrality and at the moiré band gap ($\nu = 4$), which are consistent with single particle physics, while interaction-induced peaks appear at $\nu = 1, 2, 3$. The black dashed lines trace the $R_{xx}$ peaks at integer fillings. **b**, Same as (a), but for the $\xi = 0$ device at $B_a = 50$ mT, which exhibits weaker resistance peaks that span a smaller region in $D$. **c**, Schematic cross-section of the R5LG/hBN devices. Positive $D$ pushes electron wave functions (schematically marked in purple) towards the bottom graphene layer closest to the aligned hBN and the resulting moiré. **d**, $R_{xy}$ vs. $B_a$ at $\nu = 1.37$ and $D = 0.6$ V/nm (red dots in (a) and (b)) in both devices. The $\xi = 0$ device displays anomalous Hall effect and hysteresis, whereas the $\xi = 1$ device shows no evidence of anomalous Hall resistance corresponding to the intervening nonmagnetic phase. **e**, Linecuts of $R_{xx}$ vs. $D$ along the insulating peaks at $\nu = 1, 2, 3,$ and $4$ for the two devices. The $\xi = 1$ device shows markedly higher $R_{xx}$ values that persist over a larger range of $D$.

We begin by examining the real-space structure of the moiré lattice, produced by the small lattice constant mismatch and angle $\theta$ between the graphene and hBN lattices. The corresponding moiré Brillouin zone (mBZ) is shown in Fig. 2a. As both RMG and hBN lack $C_{2z}$ symmetry, there are two inequivalent alignment orientations between the hBN and the graphene layers [9,10,13,28], corresponding to applying a π rotation to the hBN layer alone. In systems where one or both layers are $C_{2z}$ symmetric, such as twisted bilayer graphene or monolayer graphene-hBN, the two alignment orientations are equivalent. As we show next, the alignment orientation has a profound impact on the moiré R5LG.

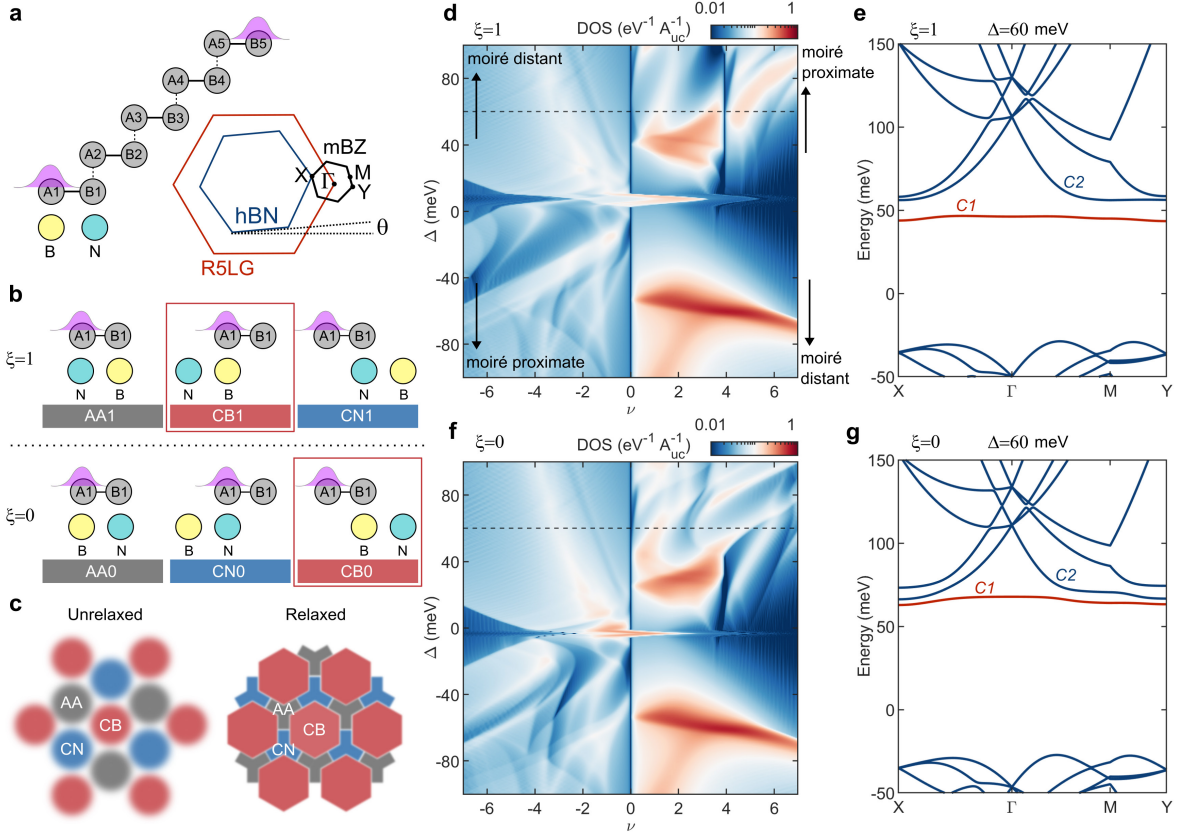

**Fig. 2. Comparison of R5LG/hBN alignment orientations $\xi$ in single-particle model. a,** Schematic side view of rhombohedral pentalayer graphene aligned to hBN and the hBN/R5LG moiré Brillouin zone (mBZ) with high-symmetry points $\Gamma, X, Y$ and $M$. The sublattice-polarized electronic wavefunctions at the top and bottom graphene layers are shown schematically in purple. **b,** Stacking configurations in the AA (gray), CB (red), and CN (blue) regions of the moiré unit cell, illustrating the relative positions of boron and nitrogen atoms with respect to graphene sublattices A1 and B1 for $\xi = 1$ (top) and $\xi = 0$ (bottom). In presence of relaxation, the CB1 and CB0 stackings (red boxes) occupy most of the moiré unit cell area as illustrated by red regions in (c). On the moiré-proximate side, the conduction electrons (purple) predominantly occupy the A1 sublattice. **c,** Real-space schematics of the moiré lattice showing that in the absence of relaxation (left) the different stacking regions are of equal size. With relaxation (right), the area of the CB stacking region expands while AA and CN regions shrink. **d,** Calculated density of states (DOS) as a function of moiré filling $\nu$ and displacement potential $\Delta$ for $\xi = 1$. **e,** Single-particle band structure of a single valley and spin for $\xi = 1$ and $\Delta = 60$ meV (dashed line in (d)), showing an isolated flat conduction band (C1). **f,** Same as (d) for $\xi = 0$. **g,** Same as (e) for $\xi = 0$. The C1 band has partial overlap with higher conduction bands.

The moiré unit cell is partitioned into three regions based on the local stacking configuration, illustrated in Figs 2b,c. In the AA regions (gray in Fig. 2c), carbon atoms from both A1 and B1 sublattices of the bottom graphene layer sit directly atop hBN atoms, while in the CB/CN regions (red/blue), atoms of a single carbon sublattice sit above either boron/nitrogen atoms, respectively. These names are analogous to the MM/MX/XM stackings of homobilayer TMD moiré systems that also lack $C_{2z}$ [33–36]. We caution that the AA/AB/BA stackings of twisted bilayer graphene behave quite differently in that AB and BA are symmetry-related, whereas CB and CN are symmetry-inequivalent in R5LG moirés.

Chemistry dictates that the most energetically favorable local stacking is the CB configuration, since the boron (nitrogen) atoms will gain a positive (negative) charge due to their electronegativity differences [8,10,11,37]. As the lattice undergoes relaxation, one expects the CB regions will thus expand, while the AA and CN regions will shrink. Indeed, both elastic theory [12,29] and *ab initio* [38] calculations find the CB region expands to occupy the majority of the moiré unit cell, especially at small angles where lattice relaxation effects are pronounced [12]. This is shown schematically in Fig. 2c, where after relaxation the CB region (red) dominates most of the unit cell. Importantly, the lattice relaxion is qualitatively identical for both alignment orientations [29].

In R5LG, an applied displacement field induces strong sublattice and layer polarization, localizing conduction band electrons of the lowest graphene layer primarily to the A1 sublattice on the moiré proximate side (Fig. 2a) [1,39]. This sublattice polarization, combined with the enlarged CB regions, produces dramatic differences between the moiré potentials of the two alignment orientations. For $\xi = 1$ (CB1 region in Fig. 2b), the A1 sites are aligned directly above boron atoms, allowing for efficient interlayer tunneling and generating a strong moiré potential. In contrast, for $\xi = 0$ (CB0 region), the A1 carbon resides over an empty site in the hBN lattice, suppressing tunnelling and significantly weakening the moiré potential. We conclude that three simple ingredients—the dominance of CB stacking due to relaxation, the A1 sublattice polarization of conduction electrons, and the alignment orientation placing boron proximate to either A1 or B1 within CB domains—thus lead to a pronounced asymmetry in moiré strength between the two orientations.

The fact that the alignment orientation selects a strong versus weak moiré potential is reflected in the single particle band structures. Modelling the electronic structure of R5LG/hBN requires specifying nearly a dozen parameters, many of which are challenging to constrain independently. To avoid over-fitting, we model the moiré potential, correctly incorporating lattice relaxation following Ref. [32] (see Methods). Lattice relaxation affects the band structure profoundly, opening gaps and shifting the positions of the van Hove singularities (Extended Data Fig. 6). This model, whose only free parameter is the alignment orientation $\xi$, was quantitatively validated in Bernal bilayer graphene-hBN experiments [32]. Figures 2e,g show the calculated band structures for $\xi = 1,0$ under a strong displacement field in the moiré-proximate regime. In both cases, the first conduction band (C1) is in the flat band regime, where strong electron correlations are expected. However, the degree of flat-band isolation differs markedly: for $\xi = 1$, the C1 band is fully separated from higher conduction bands by a sizable gap, while for $\xi = 0$, it partially overlaps with higher bands and lacks a global gap. This is a direct consequence of the dichotomy in moiré potential strengths for the two alignment orientations.

These features are reflected in the single-particle density of states (DOS) maps (Figs. 2d,f). For $\xi = 1$, a full gap above C1 persists across the moiré-proximate displacement range, as evidenced by the vanishing DOS at $\nu = 4$ in Fig. 2d (blue). In contrast, for $\xi = 0$, the analogous gap closes over most of the displacement range. Notably, both configurations exhibit a pronounced van Hove singularity near $\Delta \approx 30$–40 meV (red peaks in Figs. 2d,f), where the curvature of the C1 band switches from concave up to concave down. In the presence of interactions, such singularities promote flavor symmetry breaking and can stabilize correlated insulators. For $\xi = 1$, the well-isolated flat band supports strong insulating behavior at $\nu = 1, 2, 3,$ and $4$. In contrast, the lack of full band isolation in $\xi = 0$ weakens these tendencies, making metallic states more likely. These theoretical expectations align with the transport data in Fig. 1: the $\xi = 1$ device exhibits robust insulating states across all integer fillings, while the $\xi = 0$ device shows much weaker and more limited features. Additional pronounced differences between the two alignment orientations are observed in other transport measurements, including the Hall carrier density as shown in Extended Data Fig. 3.

**SQUID-on-tip magnetometry**

To probe the nature of the symmetry-broken states, we employ scanning SQUID-on-tip (SOT) magnetometry [40,41], which directly reveals the magnetic signatures of electronic ordering [5,42]. Spontaneous magnetization in these systems indicates broken time-reversal-symmetry (TRS), arising from spin and/or orbital moments. In spin-polarized phases, each electron contributes a fixed magnetic moment of ~1 Bohr magneton, $\mu_B$. Orbital magnetization, in contrast, stems from Berry curvature in topological bands, and thus can be significantly larger [42]. Importantly, orbital moments have opposite signs in opposite valleys, so a net orbital signal requires valley polarization.

In our measurements, an indium SOT of diameter $d \cong 200$ nm and ~10 nT/Hz$^{1/2}$ field sensitivity was scanned at a height of $h \cong 200$ nm above the sample surface in the presence of small $B_a$ (Methods). A small *ac* voltage applied to one of the gates modulates the carrier density by $n^{ac}$ ($\nu^{ac} = 4n^{ac}/n_s$), and the resulting measured *ac* magnetic signal, $B_z^{ac} = n^{ac}dB_z/dn$, is proportional to the differential magnetization $m_z = dM_z/dn$, which is the derivative of the local magnetization $M_z$ with respect to $n$.

Figure 3a shows the $B_z^{ac}$ map for the $\xi = 1$ device in the electron-doped, moiré-proximate quadrant, where correlated phases are discerned in transport. We observe $B_z^{ac}$ signals corresponding to local differential magnetization of ~5 to 15 $\mu_B$ per electron (Methods; Extended Data Fig. 4)—comparable values to magic-angle twisted bilayer graphene (MATBG) [42]—indicating TRS breaking and valley polarization. At lower $D \approx 0.4$ V/nm, strong positive (red) and negative (blue) lobes flank $\nu = 1$, consistent with a singly degenerate Fermi surface (1-fold metal), where the $K$ and $K'$ valleys of the C1 flat band are occupied sequentially on the $\nu < 1$ and $\nu > 1$ sides, respectively. Similar behavior appears around $\nu = 3$, while near $\nu = 2$ the signal is minimal, indicating a valley-symmetric ground state.

Upon increasing $D$ at $\nu = 1$, the magnetic signal is abruptly suppressed at $D \approx 0.5$ V/nm, followed by a reemergence at $D \approx 0.8$ V/nm. This reentrant behavior is unexpected from transport data, which show a single continuous peak in $R_{xx}$ at $\nu = 1$ that persists to the highest $D$ (Fig. 1a). In contrast, the $B_z^{ac}$ map (Fig. 3a) reveals a clear $D$ dependent transition into and out of an intervening weakly-magnetic (WM) phase. Figure 3b presents a schematic phase diagram for the $\xi = 1$ device, inferred from $R_{xx}$, $n_H$ (Extended Data Fig. 3), and $B_z^{ac}$ (Methods). Black vertical dashed lines mark resistive peaks in $R_{xx}$ corresponding to gapped states. Gray regions denote four-fold symmetric ground states, blue marks two-fold symmetric phases, and red indicates valley-polarized (VP) symmetry-broken states with strong magnetization, consistent with sequential filling of singly degenerate (1-fold) bands. The intervening WM phase (light blue) includes gapped states at $\nu = 1$ and $3$, and must therefore also be 1-fold. The most plausible interpretation for this 1-fold WM state (see Methods for alternatives) is an spin-polarized intervalley coherent (IVC-SP) phase, as previously observed in rhombohedral trilayer graphene [43,44].

The corresponding $B_z^{ac}$ map for the $\xi = 0$ device (Fig. 3d) reveals a qualitatively different magnetic phase diagram. At high doping, the system remains four-fold degenerate. As $\nu$ decreases, it transitions into twofold-degenerate state—likely a valley-balanced but spin ordered phase (blue region in Fig. 3e)—evidenced by a narrow negative peak in $B_z^{ac}$ along the transition line, reflecting a step in magnetization. With further reduction in $\nu$, the system enters a valley polarized, singly degenerate (1-fold) phase, characterized by strong differential magnetization near $\nu = 1$. These two consecutive transitions, from four-fold to two-fold and then to a valley-polarized 1-fold state, occur at similar $n$ and $D$ also in the moiré distant region (Extended Data Fig. 4), as well as in a non-aligned R5LG device (Fig. 3e inset; Methods). These similarities suggest that the moiré potential in the $\xi = 0$ device is relatively weak, acting as a perturbation to the unaligned system away from integer fillings. In addition, at the lowest densities, we find a highly insulating nonmagnetic phase (yellow in Fig. 3e), possibly indicating formation of a Wigner crystal (WC). Notably, the large magnetization around $\nu = 1$ in the $\xi = 0$ device persists from $D \approx 0.3$ V/nm to our highest attenable $D$. This is in contrast with $\xi = 1$ device, which displays a much weaker magnetization in the intervening WM phase at intermediate $D \approx 0.6$ V/nm values, consistent with the $R_{xy}$ data in Fig. 1d. The more complex phase diagram in the $\xi = 1$ case provides further evidence for stronger moiré effects in this alignment orientation.

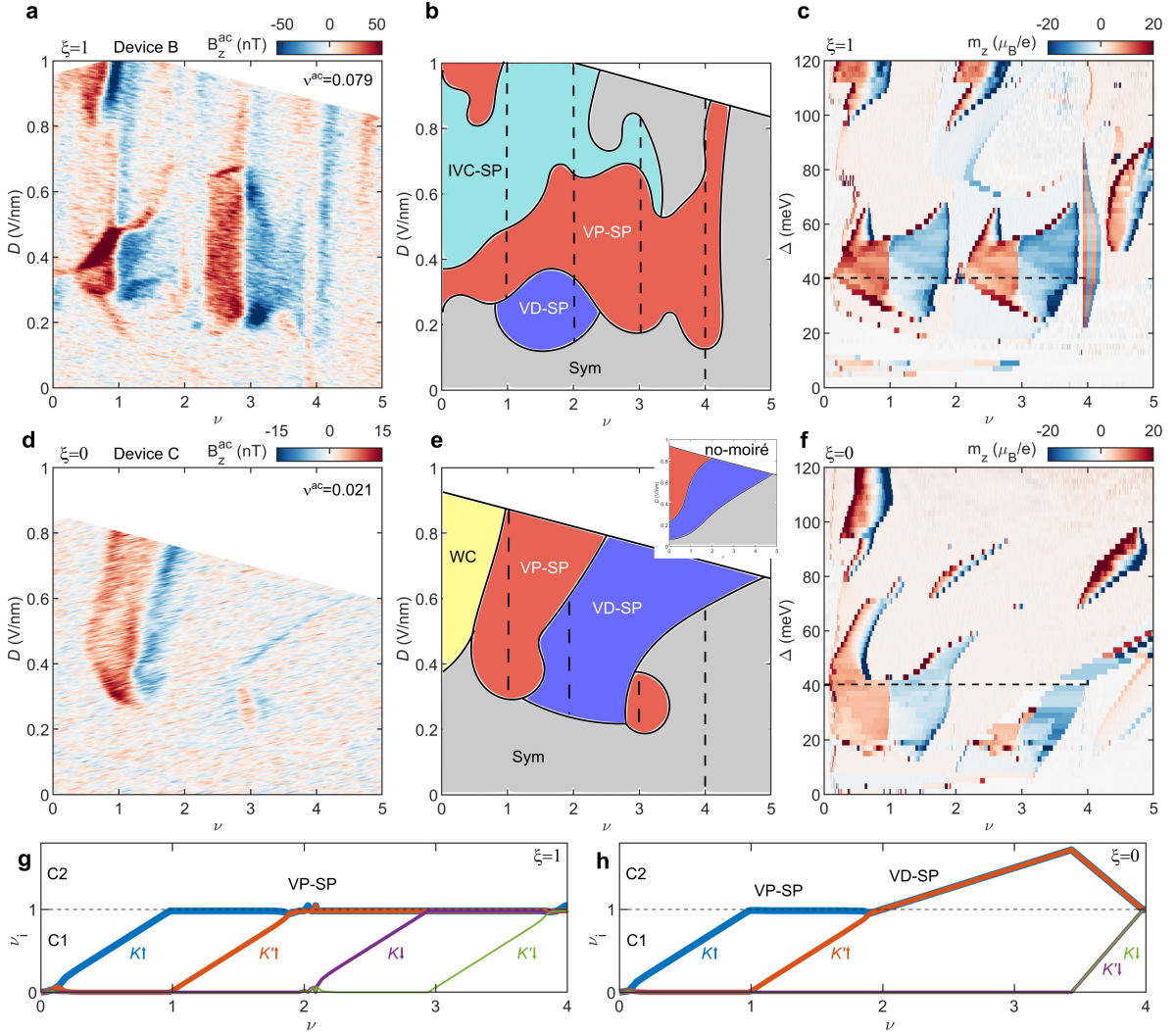

**Fig. 3. Local magnetic measurements and symmetry-broken phase diagrams. a**, $B_z^{ac}(\nu, D)$ map for the $\xi = 1$ device measured by the SOT in $B_a = 77$ mT with $\nu^{ac} = 0.079$, revealing TRS broken states with large differential magnetization. **b**, Inferred schematic phase diagram of R5LG/hBN for alignment orientation $\xi = 1$ outlining main phases: 4-fold symmetric (Sym, gray), 2-fold valley-degenerate spin-polarized (VD-SP, blue), 1-fold valley and spin polarized (VP-SP, red), and 2-fold intervening weakly-magnetic (WM) state that is spin-polarized and possibly IVC phase (IVC-SP, light blue). **c**, Differential magnetization $m_z(\nu)$ computed from the Stoner model for $\xi = 1$ (Methods). **d**, Same as (a) for the $\xi = 0$ device measured at $B_a = 10$ mT with $\nu^{ac} = 0.021$. The weaker $B_z^{ac}$ signal relative to (a) approximately scales with the relatively smaller $\nu^{ac}$. **e**, Same as (b) for the $\xi = 0$ alignment orientation. The yellow region indicates an insulating nonmagnetic phase, possibly a Wigner crystal (WC). Inset: Schematic phase diagram of R5LG without hBN alignment (see Extended Data Fig. 5). **f**, Same as (c) for $\xi = 0$. **g**, Flavor-resolved occupations $\nu_i$ in the Stoner model as a function of total filling $\nu$ for $\xi = 1$ at displacement potential $\Delta = 40$ meV along the black dashed line in (c). The different flavors fill the C1 band sequentially. **h**, Same as (g) for $\xi = 0$ alignment orientation. For $\nu > 2$, the C2 band is occupied prior to full filling of C1 band.

**Stoner model**

To gain further insight into the observed flavor-ordering transitions, we use a simple phenomenological model based on Stoner ferromagnetism. Specifically, we consider the conduction bands in the presence of strong Coulomb interactions and an effective Hund's coupling (Methods). As in the single-particle analysis, we avoid elaborate fitting and instead adopt interaction parameters previously established in studies of rhombohedral trilayer graphene [3]. Solving the model under standard mean-field approximations yields the electron density in each flavor as a function of $n$ and $D$. To compare with the SOT measurements, we compute the single-particle orbital magnetization for each flavor as a function of its density, then derive the total $M_z(\nu)$ and differential $m_z(\nu)$ magnetizations, based on the flavor-filling sequence predicted by the Stoner model (Methods).

The experimental data in Figs. 3a,d reveal a clear contrast in how the moiré minibands are filled in the $\xi = 1$ and $\xi = 0$ devices. For $\xi = 1$ at $D = 0.4$ V/nm, the system fills the C1 band sequentially, flavor by flavor, before transitioning to the second conduction band (C2). This filling sequence manifests as an alternating red–blue–red–blue $B_z^{ac}$ signal, with sign reversal at each integer fillings. Our Stoner model calculations of $m_z(\nu, D)$ (Fig. 3c) closely reproduce the key experimental features: the correct order of magnitude of $m_z$ of up to 20 $\mu_B$/e (Methods), alternating lobes of positive and negative $m_z$ near $\nu = 1$ and 3 at low $D$, a WM phase at intermediate $D$, and reentrant magnetization at high $D$. Figure 3g further supports this interpretation, showing the calculated flavor-resolved occupations $\nu_i$ as a function of total filling $\nu$ at displacement potential $\Delta = 40$ meV. The model predicts that the system sequentially fills the C1 band in the order $K\uparrow$, $K'\uparrow$, $K\downarrow$, and $K'\downarrow$ before any occupation of C2 begins, consistent with experimental observations.

The Stoner calculations for $\xi = 0$ yield a sharply different phase diagram (Fig. 3f) that also qualitatively matches the SOT data in Fig. 3d. The model captures the alternating $m_z$ near $\nu = 1$, the weak magnetic signal in the spin-polarized phase for $\nu > 2$, and the negative $m_z$ associated with transition to a fully symmetric metallic state for $\nu > 3$. The flavor filling sequence, however, differs markedly from the $\xi = 1$ case. In particular, the red lobe in Fig. 3d persists beyond $\nu = 1$, suggesting that after one flavor of the C1 band is filled, the system then fills another band in the same valley, likely C2. This interpretation is supported by the Stoner results in Fig. 3h, which show that C2 begins to fill above $\nu = 2$, before all four flavors in C1 are fully occupied.

The contrasting behavior of the two alignment orientations can be traced back to the close competition between exchange energy, which favors flavor polarization, and the single-particle gap to the second conduction band. For $\xi = 1$, the presence of a sizable gap above C1 favors full occupation of each flavor within C1 before accessing higher bands, resulting in sequential flavor filling. For $\xi = 0$, by contrast, the absence of a well-defined gap makes it energetically favorable to begin filling C2 before C1, leading to a different hierarchy of symmetry breaking. These results highlight the importance of band isolation in shaping the interaction-driven ground states. Remarkably, this simple Stoner model—without any fitting parameters—captures the distinct phase diagrams and flavor filling sequences observed in both $\xi$ configurations.

**Discussion**

Our results demonstrate that the hBN alignment orientation $\xi$ is a crucial determinant of correlated phase behavior in rhombohedral graphene/hBN moiré systems. Devices with nearly identical twist angles but different alignment orientation $\xi$ exhibit strikingly distinct sequences of symmetry-broken states, magnetic transitions, and miniband structures—particularly on the moiré-proximate side, where moiré effects are strongest. For $\xi = 1$, R5LG/hBN displays Stoner magnetism and strongly correlated flat-band physics, whereas for $\xi = 0$, the system more closely resembles non-moiré R5LG, exhibiting only weak signatures of a moiré-induced effects.

A minimal Stoner ferromagnetism model reproduces the key experimental features of both device types, including the number and type of magnetic lobes, the hierarchy of flavor symmetry breaking, and the reentrant magnetization in the $\xi = 1$ case. Crucially, all parameters of the model except $\xi$ are held fixed. This shows that a single theoretical framework—without fitting—can account for the distinct behaviors of both $\xi$ configurations. The effect of $\xi$ arises from the combined influence of sublattice polarization, moiré relaxation, and the local stacking configuration between graphene and hBN: depending on the alignment orientation, the sublattice-polarized conduction electrons experience enhanced or suppressed carbon-boron tunneling, resulting in a pronounced asymmetry in moiré potential strength.

Currently, there is no experimental technique capable of determining $\xi$ during the assembly process. Consequently, devices cannot be systematically fabricated with a known alignment orientation, and nominally identical stacking protocols may result in different $\xi$ values. This fundamental uncertainty poses a challenge for reproducibility and complicates the interpretation of experimental results across devices. In addition, the layered structure of hBN—where each successive atomic plane is rotated by 180°—means that even a single atomic step in the hBN substrate locally flips the alignment orientation. Such steps can introduce sharp transitions in $\xi$ within the same device, leading to spatial inhomogeneities and complex global behavior. Together, these two effects provide a natural explanation for the variability and apparent inconsistencies observed in moiré graphene systems.

Our new understanding of sublattice alignment orientation offers a powerful framework to retrospectively categorize earlier studies that overlooked this structural parameter. In particular, strong insulating behavior at $\nu = 3$ and 4 on the moiré-proximate side serves as a robust empirical signature of $\xi = 1$ devices, enabling reliable assignment of alignment orientation based on electron-side transport phase diagrams. As summarized in Table 1, this approach allows for consistent classification of nearly all published rhombohedral moiré graphene devices spanning four to seven layers, helping to resolve apparent discrepancies between seemingly similar systems.

| Reference | No of layers | Device | Moiré wavelength/Angle | $\xi$ |
|---|---|---|---|---|
| Choi *et al.*, [45] | 4 | A | 14.5 ~14.8 nm | 0 |
| Lu *et al.*, [18] | 5 | 1 | 11.5 nm, 0.77° | 1 |
| Waters *et al.*, [16] | 5 | - | 10.8 nm, 0.90° | 0 |
| Aronson *et al.*, [17] | 5 | D1, D2 | 12.4 nm, 0.63° | 0 |
| Li *et al.*, [46] | 5 | - | 10.1 nm, 1.02° | 0 |
| This work | 5 | A | 14.69 nm, 0.06° | 1 |
| This work | 5 | B | 14.20 nm, 0.26° | 1 |
| This work | 5 | C | 14.05 nm, 0.29° | 0 |
| This work | 5 | D | 14.05 nm, 0.29° | 0 |
| Xie *et al.*, [19] | 6 | - | 14.5 nm, 0.17° | 0 |
| Zhou *et al.*, * [47] | 7 | D2 | Top hBN 0.88° | 0 |
|  |  |  | Bottom hBN 0.90° | 1 |
| Ding *et al.*, [48] | 7 | D1 | 0.86° | 1 |
| Wang *et al.*, [20] | 7 | - | 12.3 nm | 0 |

**Table 1**. **Classification of putative alignment orientation $\xi$ for previously reported RMG/hBN devices.** The devices are ordered by the number of graphene layers. Nine RMG/hBN interfaces show $\xi = 0$, while only five display $\xi = 1$. *This work reported a doubly-aligned rhombohedral heptalayer graphene with different top and bottom $\xi$ orientations according to our classification.

Notably, the resulting trend indicates that $\xi = 1$ is less frequently realized in experiments—possibly because its stronger moiré potential increases susceptibility to relaxation towards Bernal stacking, misalignment, or other forms of disorder that degrade device quality. The alignment orientation determines the type of strongly correlated topological phases that appear. Interestingly, Chern insulators have only been observed on the moiré proximate side in samples whose putative alignment orientation is $\xi = 0$ [17]. This asymmetry between alignment orientations thus manifests not only in energetic differences, but also in topological properties. Another important observation is that both alignment orientations appear capable of supporting integer and fractional quantum anomalous Hall states at $\nu = 1$ and $\nu = 2/3$ on the moiré-distant side [18,19,49], where the influence of sublattice alignment is comparatively weaker. Still, because these phases originate from the hBN-induced moiré potential, their presence and stability may be subtly—but decisively—influenced by $\xi$. For instance, alignment orientation could shift the twist angle range over which these states are realized, or affect which fractional fillings are stabilized. This may help explain why to date certain fractional Chern insulator states have been observed conclusively in only a single work [18], suggesting that $\xi$ could play a critical and previously unrecognized role in accessing fragile topological phases in moiré systems.

This framework may also help clarify a longstanding puzzle in MATBG aligned to hBN: the quantum anomalous Hall effect has been observed only rarely, and often only in a specific region of the device [49,50]. It is possible that MATBG aligned to hBN undergoes lattice relaxation that breaks $C_{2z}$ and, therefore, sublattice alignment plays a similarly decisive role in MATBG, with the QAHE stabilized only when $\xi = 1$. If $\xi = 1$ is indeed rare—as our analysis suggests—this could help explain both the extreme scarcity and the apparent irreproducibility of the QAHE in that system.

Finally, our findings underscore the need for theoretical models to explicitly include $\xi$ as a structural input. The assumption that $\xi$ averages out due to moiré symmetry is invalid in $C_{2z}$-breaking systems like RMG, where sublattice- and layer-polarization amplify sensitivity to atomic-scale orientation. The alignment $\xi$ is not a small perturbation — it reshapes the entire interaction landscape and determines the nature of the emergent correlated ground states. Identifying and controlling $\xi$ is therefore essential for understanding and engineering the full range of symmetry-breaking phenomena in moiré systems.

**Acknowledgments**

J.D. and D.E.P. acknowledge discussions with T. Soejima. This work was co-funded by the Minerva Foundation grant No 140687, by the United States - Israel Binational Science Foundation (BSF) grant No 2022013, by the Institute for Artificial Intelligence, and by the European Union (ERC, MoireMultiProbe - 101089714). Views and opinions expressed are however those of the author(s) only and do not necessarily reflect those of the European Union or the European Research Council. Neither the European Union nor the granting authority can be held responsible for them. E.Z. acknowledges the support of the Goldfield Family Charitable Trust, Leona M. and Harry B. Helmsley Charitable Trust grant #2112-04911, the MIT-Israel Zuckerman STEM Fund, and grant no. NSF PHY-2309135 to the Kavli Institute for Theoretical Physics (KITP). J.D. was funded by NSF DMR-2220703. K.W. and T.T. acknowledge support from the JSPS KAKENHI (Grant Numbers 21H05233 and 23H02052), the CREST (JPMJCR24A5), JST and World Premier International Research Center Initiative (WPI), MEXT, Japan. This work used the Expanse cluster at the San Diego Supercomputer Center through allocation PHY240272 from the Advanced Cyberinfrastructure Coordination Ecosystem: Services & Support (ACCESS) program, which is supported by U.S. National Science Foundation grants #2138259, #2138286, #2138307, #2137603, and #2138296.

## Author contributions

M.U. fabricated and characterized the samples. M.U., W.Z., S.D. and N.A. performed the local magnetization studies. M.U., W.Z., N.A., and S.D. performed the transport measurements and data analysis. M.U. and E.Z. designed the experiment. W.Z., S.D., M.L., N.S.K. and Y.M. fabricated the SOT and the tuning fork, and M.E.H. developed the SOT readout. M.U., M.B., D.E.P. and J.D. performed the single-particle theoretical modeling. J.D. and D.E.P. performed the Stoner calculations. K.W. and T.T. provided the hBN crystals. M.U., M.B., D.E.P., J.D., W.Z. and E.Z. wrote the original manuscript. All authors participated in discussions and revisions of the manuscript.

**Competing interests** The authors declare no competing interests.

## Methods

### Device fabrication

The hBN-encapsulated pentalayer graphene heterostructures were fabricated using a dry-transfer method optimized for rhombohedral multilayer graphene, as described in detail in [5]. To distinguish rhombohedral from Bernal-stacked regions, we mapped the Raman G peak position both before and after stacking [51], as shown in Extended Data Fig. 1. During assembly, we deliberately aligned the crystal axes of the pentalayer graphene with one of the hBN flakes using a mechanical rotation stage. Naturally occurring straight edges on each flake served as visual guides. To avoid unintentional 30° misalignment, we determined the edge chiralities (zigzag or armchair) of both the hBN and graphene by analyzing the polarization dependence of second harmonic generation (SHG) in the hBN flakes and in a Bernal-stacked trilayer region on the same graphene crystal [52,53].

Device summary:

**Devices A and B ($\xi = 1$)**: Ti (2 nm) / Au (10 nm) top gate and Ti (2 nm) / Pt (10 nm) bottom gate, the top and bottom hBN thicknesses are ≈16.8 nm and 36.6 nm, respectively. $n_s$ ($\nu = 4$) for Device A was $2.14 \cdot 10^{12}$ cm$^{-2}$ corresponding to alignment angle $\theta \approx 0.06°$, and $2.29 \cdot 10^{12}$ cm$^{-2}$ and $\theta \approx 0.26°$ for Device B.

**Devices C and D ($\xi = 0$)**: Ti (2 nm) / Au (10 nm) top gate and graphite bottom gate, with top and bottom hBN thicknesses of ≈25.3 nm and 27 nm, respectively. $n_s$ ($\nu = 4$) for the two devices was $2.33 \cdot 10^{12}$ cm$^{-2}$ corresponding to $\theta \approx 0.29°$.

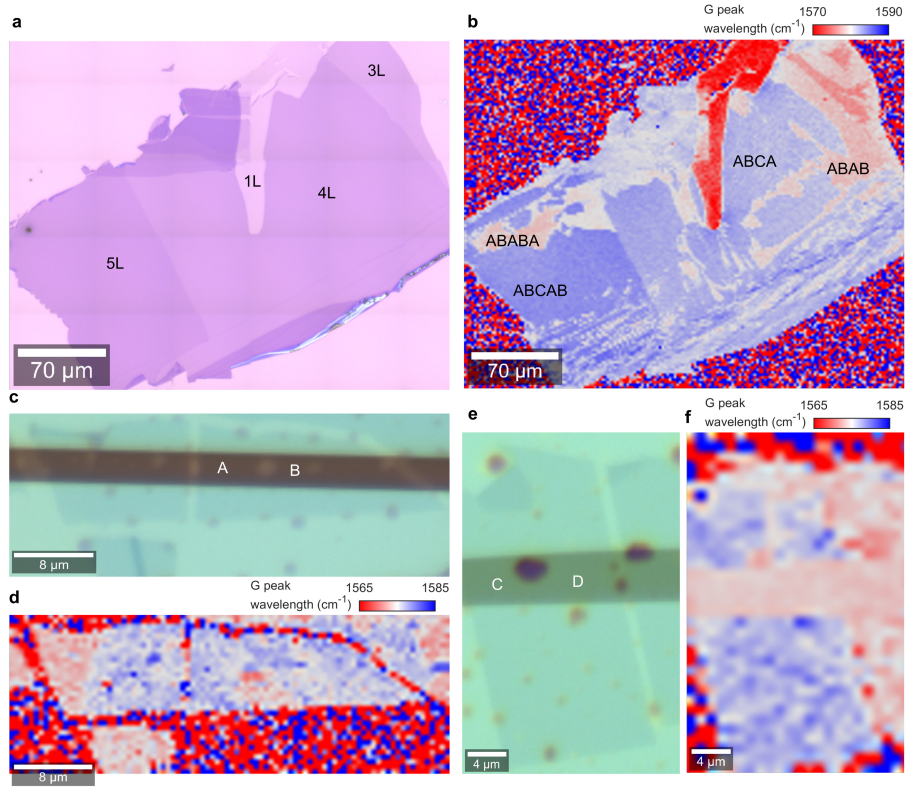

**Extended Data Fig. 1. Device fabrication. a.** Optical image of the initial graphite flake on Si/SiO$_2$ substrate for all the devices. The different number of graphene layers 1L, 3L, 4L, 5L in the various regions are presented by the different shades of purple. **b.** Raman map of the G peak wavelength of the flake in (a). The different stacking regions are marked. Rhombohedral stacking shifts the G peak to lower wavelengths compared to Bernal. **c.** Optical image of the final stack for devices A and B. **d.** Same as (b) for the final stack in (c). **e.** Same as (c) for devices C and D. **f.** Same as (d) for the final stack in (e).

**Transport measurements of R5LG/hBN**

The transport measurements were performed using standard lock-in techniques and low-noise voltage amplifiers with $ac$ bias current $I^{ac}$ from 3 nA to 10 nA at $f = 13.777$ Hz under cryogenic temperatures ranging from 300 mK to 4 K. We applied $dc$ voltages to the top and bottom gates, $V_{tg}^{dc}$ and $V_{bg}^{dc}$, to set the carrier density $n = (C_{tg}V_{tg}^{dc} + C_{bg}V_{bg}^{dc})/e$ and the displacement field $D = (C_{tg}V_{tg}^{dc} - C_{bg}V_{bg}^{dc})/2\varepsilon_0$, where $C_{tg}$ and $C_{bg}$ are the top-gate and bottom gate capacitance per unit area of the top and bottom gates, respectively. Extended Data Table 1 summarizes all the conditions for different transport measurements.

| Figure | Temperature | Current $I^{ac}$ (nA) | Conditions |
|---|---|---|---|
| Fig. 1a | 500 mK | 10 | $B_a = 0$ T |
| Fig. 1b | 500 mK | 3 | $B_a = 50$ mT |
| Fig. 1d $\xi = 1$ | 500 mK | 10 | $\nu = 1.37, D = 0.6$ V/nm |
| Fig. 1d $\xi = 0$ | 300 mK | 10 | $\nu = 1.37, D = 0.6$ V/nm |
| ED Fig. 2a, d | 4 K | 10 | $B_a = 0$ T |
| ED Fig. 2b | 500 mK | 10 | $B_a = 0$ T |
| ED Fig. 2c | 1 K | 10 | $B_a = 0$ T |
| ED Fig. 3a | 4 K | 10 | $D = 0$ V/nm |
| ED Fig. 3b | 500 mK | 10 | $B_a = \pm 0.3$ T |
| ED Fig. 3c | 300 mK | 10 | $B_a = \pm 0.15$ T |
| ED Fig. 5a | 500 mK | 10 | $B_a = 0$ T |
| ED Fig. 5b | 500 mK | 5 | $B_a = 3$ T |
| ED Fig. 5c | 500 mK | 10 | $D = 0$ V/nm |
| ED Fig. 5e | 500 mK | 5 | $D = 0.85$ V/nm |

**Extended Data Table 1**. Summary of transport measurements conditions presented in the main text and Methods.

Extended Data Fig. 2 summarizes $R_{xx}(\nu, D)$ at $B_a = 0$ T of all four studied devices. The overall $R_{xx}$ maps are consistent across devices with the same $\xi$, demonstrating reproducibility of the observed symmetry-broken states. All devices exhibit similar behavior at $|D| > 0.2$ V/nm on the moiré distant side, but show pronounced differences on the moiré proximate side. Device A ($\xi = 1$) shows well-developed resistance peaks at all integer fillings ($\nu = 1,2,3,4$) even at $T = 4$ K.

To determine the twist angle, we measure Landau fan diagram $R_{xx}\left(\nu, \frac{\phi}{\phi_0}\right)$, where $\phi = B_a A = 4B_a/n_s$ is the magnetic flux in moiré unit cell area $A$. Extended Data Fig. 3a presets $R_{xx}\left(\nu, \frac{\phi}{\phi_0}\right)$ at $D = 0$ V/nm in device D along with the fit to the Brown Zak oscillations (dashed lines). We extract $A$ from the fit and calculate the moiré wavelength $\lambda$ and the twist angle $\theta$:

$$\lambda = \sqrt{\frac{2A}{\sqrt{3}}} = \frac{(1+\delta)a_0}{\sqrt{2(1+\delta)(1-\cos\theta) + \delta^2}} \qquad (1)$$

where $a_0 = 0.246$ nm is the graphene lattice constant and $\delta = \frac{a_{BN}}{a_0} - 1 = 0.017$ is the lattice mismatch between graphene and hBN.

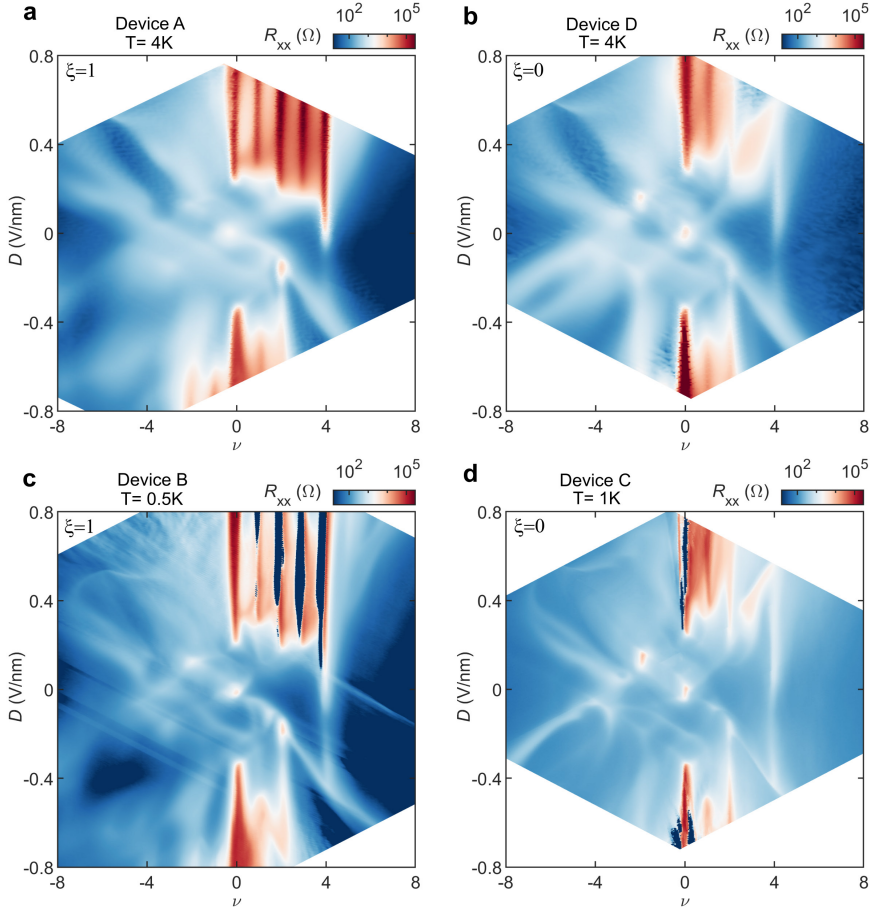

**Extended Data Fig. 2**. **Comparison of $R_{xx}$ maps of four R5LG/hBN devices. a.** $R_{xx}(\nu, D)$ map of device A ($\xi = 1$) measured at $T = 4$ K similar to the data in Fig. 1a measured at $T = 0.5$ K. **b.** Same as (a) for device D ($\xi = 0$), similar to the data in Fig. 1b. **c.** $R_{xx}(\nu, D)$ map of device B ($\xi = 1$), used for local magnetometry study in Fig. 3a. **d.** Same as (c) for device C ($\xi = 0$).

Extended Data Figs. 3b,c show the Hall density maps, $n_H = \left[e\left(\frac{\delta R_{xy}}{\delta B_a}\right)\right]^{-1}$, for the two $\xi$ samples on the electron-doped, moiré-proximate side. In these maps, zero-crossings of $n_H$ indicate band gaps, while divergences with abrupt sign changes mark van Hove singularities (vHs) (panel c inset). We highlight these features with black solid lines (gaps) and white dashed lines (vHs). The number and position of these features provide information about the degeneracy structure of the bands: in the symmetric case (fourfold degeneracy), one expects no gaps and one vHs between $\nu = 0$ and 4; in a twofold-degenerate system, one expects one gap and two vHs; and in a fully flavor-polarized (onefold) case, three gaps and four vHs should appear. These expectations are consistent with the data. In the $\xi = 1$ device A, only a single vHs is observed at $D < 0.2$ V/nm. As the $\nu = 2$ gap opens at $D \approx 0.2$ V/nm, the vHs splits into two, and further splits into four when the $\nu = 1$ and 3 gaps appear—consistent with a transition into a onefold symmetry-broken state.

In contrast, the $\xi = 0$ device D exhibits a different sequence. A vHs initially appears between $\nu = 2$ and 3 in the symmetric phase. As $D$ increases, it abruptly shifts to lie between $\nu = 1$ and 2, coinciding with the opening of the $\nu = 2$ gap. It then shifts again to lie between $\nu = 0$ and 1 as the onefold valley-polarized phase emerges. This "replica" evolution of vHs and gaps—where the singularity migrates in discrete steps as symmetry-breaking progresses—is reminiscent of behavior seen in non-aligned rhombohedral graphene, where the moiré potential only opens gaps in phases with different spin and valley degeneracies.

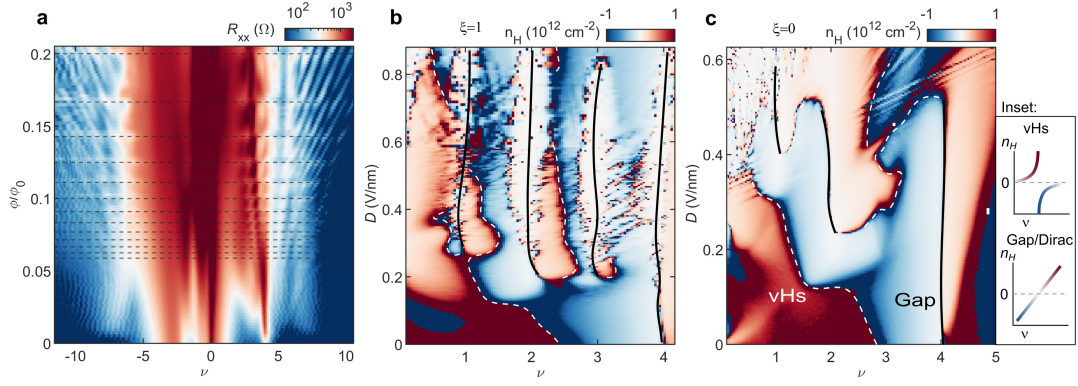

**Extended Data Fig. 3. Additional transport data on R5LG/hBN. a.** Landau fan diagram of $R_{xx}(\nu, \phi/\phi_0)$ where $\phi = B_a A$ and $A = 4/n_s$, measured in device D at $D = 0$ V/nm. Positions of the horizontal resistance features (dashed lines) are fitted to rational flux values $\frac{\phi}{\phi_0} = 1/q$, where $q$ is an integer, to extract the moiré unit cell area $A$. **b.** Hall carrier density map $n_H(\nu, D)$ at $B_a = \pm 0.3$ T for the $\xi = 1$ device (device A). **c.** Same as (b), for the $\xi = 0$ device (device D), measured at $B_a = \pm 0.15$ T. Inset: cartoon of the $n_H$ evolution across a van Hove singularity (vHs dashed white lines in (b) and (c)) and across a gap or Dirac point (solid black lines in (b) and (c)).

**Local magnetic imaging**

Indium SOTs were fabricated as described previously [54]. Local magnetic imaging for devices A and B ($\xi = 1$) were performed in wet dilution refrigerator at $T = 20$ mK, while magnetic imaging of device C ($\xi = 0$) was carried out in a Helium-3 refrigerator at $T = 300$ mK. The SOT signal was read out using a cryogenic SQUID array amplifier (SSAA) [55,56]. To maintain the scanning height ≈200 nm above the surface of the sample, the SOT was mounted on a quartz tuning fork (Model TB38, HMI Frequency Technology) vibrating at its resonance frequency ~33 kHz which acts as a force sensor [57]. *dc* voltages where applied to the top and bottom gates, $V_{tg}^{dc}$ and $V_{bg}^{dc}$, to set $n$ and $D$, along with a small *ac* modulation to one of the gates, $V_{tg}^{ac}$ or $V_{bg}^{ac}$, at frequencies of about 1 kHz. This modulates the carrier density and displacement field: $n^{ac} = (C_{tg}V_{tg}^{ac} + C_{bg}V_{bg}^{ac})/e$, $D^{ac} = (C_{tg}V_{tg}^{ac} - C_{bg}V_{bg}^{ac})/2\varepsilon_0$, and the corresponding *ac* magnetic field $B_z^{ac}$ is measured by the SOT using a lock-in amplifier.

Single-point local magnetic measurements shown in Figs. 3a,d were performed by fixing the spatial position of the SOT, varying the *dc* voltages applied to the gates $V_{tg}^{dc}$ and $V_{bg}^{dc}$, and measuring the corresponding $B_z^{ac}$. Figure 3a and Extended Data Figs. 4a-c were measured at $B_a = 77$ mT, $V_{bg}^{ac} = 100$ mVrms ($n^{ac} = 4.53 \cdot 10^{10}$ cm$^{-2}$) and Fig. 3d and Extended Data Fig. 4b were measured at $B_a = 10$ mT, $V_{bg}^{ac} = 20$ mVrms ($n^{ac} = 1.23 \cdot 10^{10}$ cm$^{-2}$).

Extended Data Figs. 4d,e show two-dimensional $B_z^{ac}(x, y)$ map and corresponding reconstruction of the differential magnetization $m_z(x, y)$ using a numerical inversion algorithm where instead of the Tikhonov regularized Fourier transform-based approach [58], we employ a numerical method based on a deep neural network, similar to [59]. The magnitude of the attained $m_z^{ac} = dM_z/dn = M_z^{ac}/n^{ac}$ of up to 70 µB/e reflects a large magnetization change across a magnetic transition (magenta circle in Extended Data Figs. 4b), whereas in other magnetically ordered states in the phase diagram the $m_z^{ac}$ is ~5 to 15 µB/e, consistent with orbital magnetization. This procedure forms the basis for identifying magnetic regions and constructing the schematic phase diagrams shown in Fig. 3b and 3e.

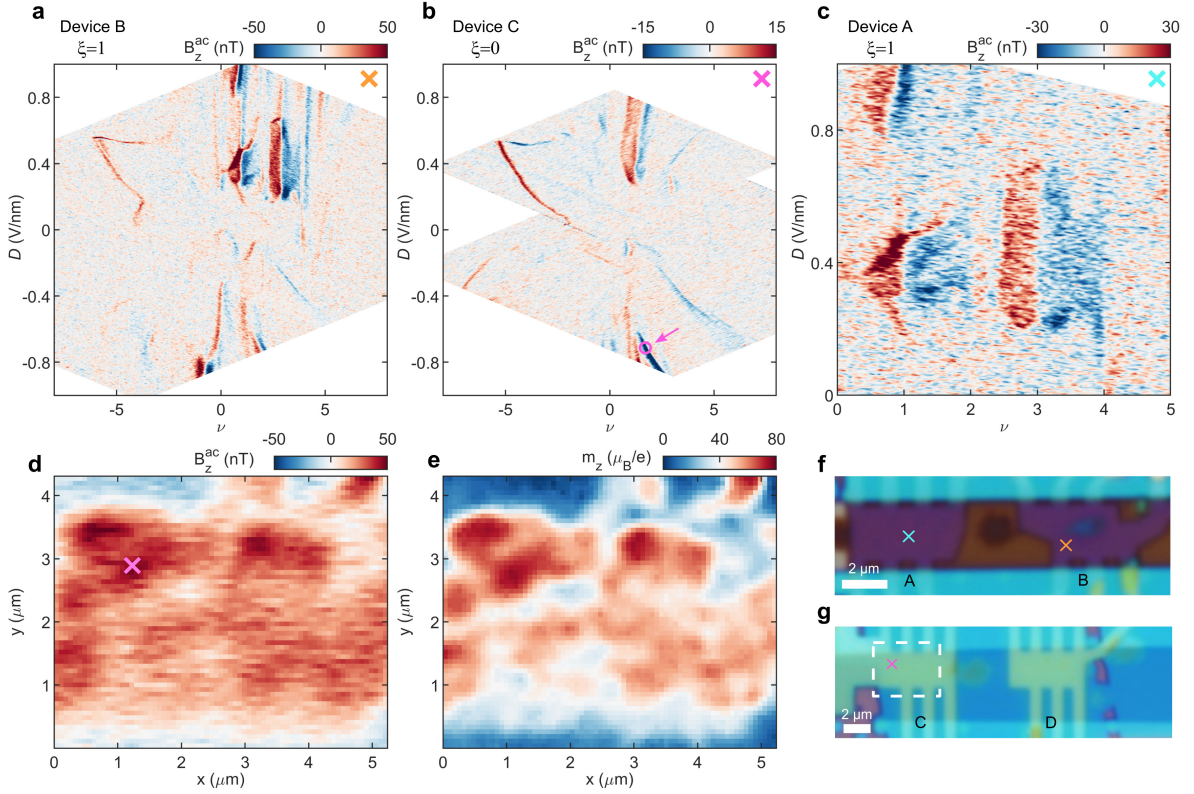

**Extended Data Fig. 4. Additional local magnetization data. a.** Full range $B_z^{ac}(\nu, D)$ map in $\xi = 1$ device B acquired at the location marked by orange cross in (f). **b.** $B_z^{ac}(\nu, D)$ map in $\xi = 0$ device C acquired at the location marked by magenta crosses in (d) and (g). **c.** $B_z^{ac}(\nu, D)$ in $\xi = 1$ device A acquired at the cyan cross in (f). **d.** Spatial $B_z^{ac}(x, y)$ map of device B in the area marked by dashed rectangle in (g) measured at $(\nu, D)$ marked by the magenta circle in (b). **e.** Differential magnetization $m_z(x, y)$ numerically reconstructed from (d). **f,g,** Optical images of devices A and B (**f**) and C and D (**g**).

**Transport measurements on a non-aligned rhombohedral pentalayer graphene device**

To investigate the effect of the moiré superlattice potential, we also performed transport measurements on a non-aligned R5LG device. Extended Data Fig. 5a shows $R_{xx}(n, D)$ measured at $B_a = 0$ T. At $n = 0$, we observe correlated insulator from $D = 0$ to 0.1 V/nm and band insulator at $D > 0.2$ V/nm. Additionally, the $R_{xx}$ map is separated into a few regions by weak resistance peaks. To unravel the possible symmetry breaking phases, we acquire $R_{xx}(n, D)$ and $R_{xy}(n, D)$ at $B_a = 3$ T, and derive the differential conductance $d\sigma_{xy}/dn$ (Extended Data Fig. 5b), where $\sigma_{xy} = \frac{R_{xy}}{R_{xx}^2 + R_{xy}^2}$ is the transverse conductance. Shubnikov-de Haas oscillations with different frequencies are clearly seen in different regions in the $n$-$D$ plane. By measuring Landau fan diagram at $D = 0$ V/nm (Extended Data Fig. 5c) and $D = 0.85$ V/nm (Extended Data Fig. 5e), we perform Fourier transform of $R_{xx}(1/B_a)$ (Extended Data Figs. 5d and f). At $D = 0$ V/nm, the Fermi surface is simply connected and four-fold degenerate, with FFT peaks at $\frac{f}{n} = 0.25\ h/e$, for $n > 1 \cdot 10^{12}$ cm$^{-2}$. In contrast, at $D = 0.85$ V/nm, the Fermi surface exhibits a 1-fold degenerate quarter-metal state ($\frac{f}{n} = 1\ h/e$) for $n < 1.5 \cdot 10^{12}$ cm$^{-2}$, and transitions to a 2-fold degenerate half-metal state ($\frac{f}{n} = 0.5\ h/e$) at higher densities. The corresponding phase diagram is sketched schematically in Fig. 3e inset and is similar to the one obtained in [6]. Compared to the hBN-aligned devices, the $\xi = 0$ devices show a similar symmetry breaking sequence on the moiré proximate side, indicating a perturbative role of the superlattice potential, while the $\xi = 1$ devices show a much stronger effect with fundamentally different phases.

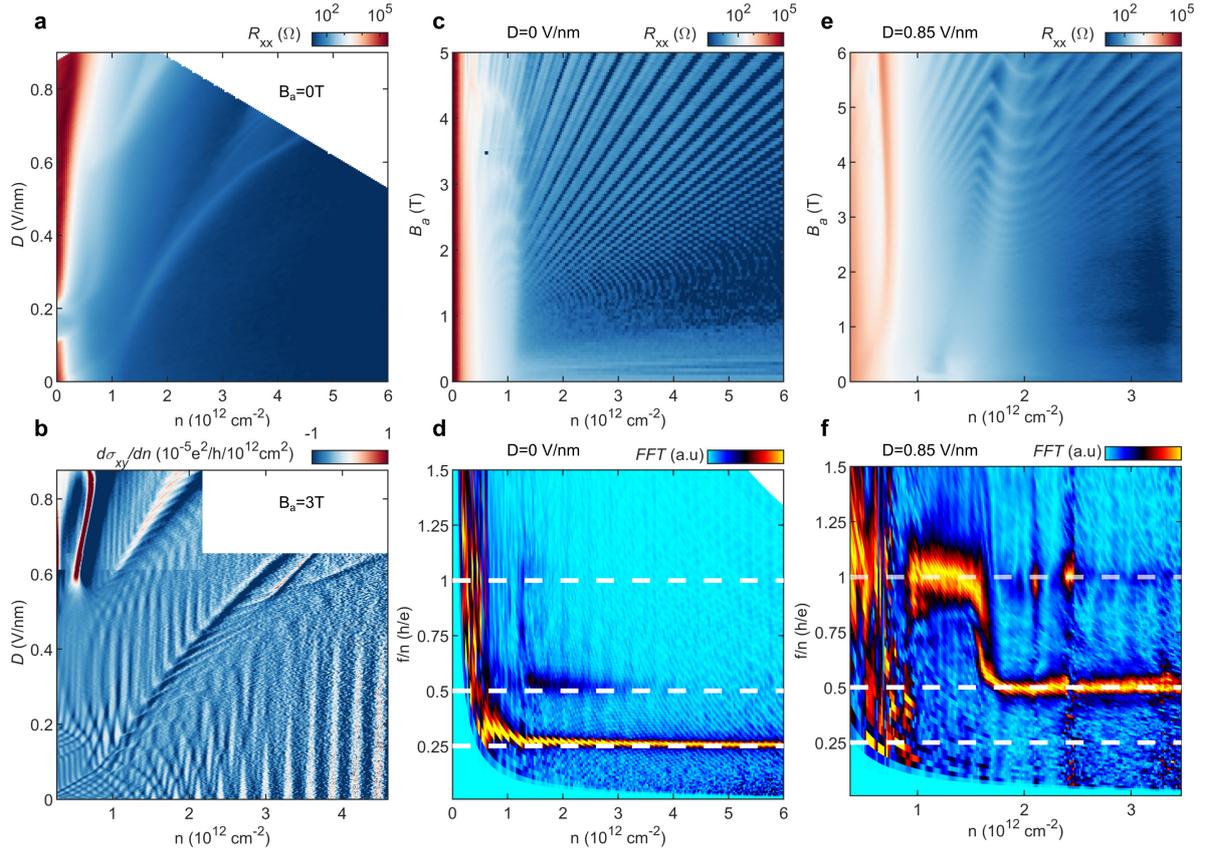

**Extended Data Fig. 5. Transport data on non-aligned R5LG**. **a.** $R_{xx}(n,D)$ map at $B_a = 0$ T. **b.** $\frac{d\sigma_{xy}}{dn}(n,D)$ map at $B_a = 3$ T, showing Shubnikov-de Haas oscillations. **c.** Landau fan diagram of $R_{xx}(n, B_a)$ at $D = 0$ V/nm. **d.** Fast Fourier transform (FFT) spectrum of $R_{xx}(1/B_a)$ as a function of $n$, with normalized frequency $f/n$ shown in units of $h/e$, calculated from the data in (c) over $B_a$ range $[1\,T, 4\,T]$. **e.** Same as (c) at $D = 0.85$ V/nm. **f.** Same as (d) calculated from the data in (e) over the $B_a$ range $[1\,T, 6\,T]$.

**Continuum model**

Modelling RMG/hBN requires nearly a dozen parameters that are difficult to constrain independently. To avoid over-fitting, we use models whose quantitative validity has been established in experimental studies of simpler rhombohedral systems. Explicitly, we model RMG using tight-binding models from non-moiré R4LG [2,5], hBN tunneling parameters from the weakly interacting moiré system R2LG/hBN [32], and Stoner models validated in non-moiré R3LG [3]. To model M-layer Rhombohedral graphene [1,60] we use models from [2,5] (see also [3,13,61,62]). We use a graphene lattice $\boldsymbol{a_1} = (1,0)a$, $\boldsymbol{a_2} = (1/2, \sqrt{3}/2)a$, with $a \approx 2.46\,\text{Å}$. We consider orbitals at $\boldsymbol{r}_{A,l} = \left(0, \frac{l-1}{\sqrt{3}}a, l_c\right), \boldsymbol{r}_{B,l} = \left(0, \frac{l}{\sqrt{3}}a, l_c\right)$, where $l_c \approx 3.4\,\text{Å}$ is the interlayer spacing. We use an orbital basis $|\sigma, l\rangle$ with sublattice $\sigma = \{A, B\}$ and layer $l = \{1, \dots, 5\}$. The corresponding tight-binding model is a block matrix in layer space

$$h^{R5G} = \begin{bmatrix} T_0 & T_1 & T_2 & 0 & 0 \\ T_1^\dagger & T_0 & T_1 & T_2 & 0 \\ T_2^\dagger & T_1^\dagger & T_0 & T_1 & T_2 \\ 0 & T_2^\dagger & T_1^\dagger & T_0 & T_1 \\ 0 & 0 & T_2^\dagger & T_1^\dagger & T_0 \end{bmatrix} + h^\Delta, \quad (2)$$

where the T's are matrices in sublattice space

$$T_0(k) = \begin{pmatrix} 0 & -t_0 f_{\boldsymbol{k}} \\ -t_0 \bar{f}_{\boldsymbol{k}} & 0 \end{pmatrix}, T_1(k) = \begin{pmatrix} -t_4 f_{\boldsymbol{k}} & -t_3 \bar{f}_{\boldsymbol{k}} \\ t_1 & -t_4 f_{\boldsymbol{k}} \end{pmatrix}, T_2(k) = \begin{pmatrix} 0 & \frac{t_2}{2} \\ 0 & 0 \end{pmatrix}, f_{\boldsymbol{k}} = \sum_{n=0}^{2} e^{i \boldsymbol{k} \cdot \delta_n}. \quad (3)$$

Here we used nearest neighbor vectors $\delta_n = R_{n2\pi/3}(0, \frac{1}{\sqrt{3}}a)$ for $n = 0, 1, 2$ with $R_\theta$ the counterclockwise rotation matrix by angle $\theta$. Finally, the displacement field and interlayer potentials are modelled by a diagonal matrix

$$h^\Delta_{\sigma l, \sigma' l'} = \delta_{\sigma l, \sigma' l'}[\mathbf{D}_l + \delta_1(1 - \delta_{\sigma l, A_1} - \delta_{\sigma l, B_5})], \quad (4)$$

$$\mathbf{D}_l = \Delta(1, 1/2, 0, -1/2, -1) + \delta_2(1, -1, -1, -1, 1). \quad (5)$$

The $\delta_1$ term captures a surface charge that forms on the "unpaired" $A_1$ and $B_5$ sites and $\delta_2$ is an constant internal screening potential [2] (see [15] for an alternative approach). See Extended Data Table 2 for parameter values.

The lattice for hBN is $\tilde{\boldsymbol{a}}_i = MR_\theta \boldsymbol{a}_i$ where $M = (1 + \epsilon)I$ is a dilation where $\epsilon \approx 0.017$ and $R_\theta$ is the counterclockwise rotation matrix by twist angle $\theta$. In the vicinity of its K-point, hBN is modelled by a constant Hamiltonian $h^{hBN} = diag\ (V_B, V_N)$ in the sublattice basis. We take $V_B = 3.34\ eV$ and $V_N = -1.4\ eV$ [9].

The RMG/hBN moiré lattice is $\boldsymbol{R}_i = (I - R^{-1}M^{-1})\boldsymbol{a}_i$, with period $L_M$ and angle $\phi$ between $\boldsymbol{R}_1$ and $\boldsymbol{a}_1$ of

$$L_M = \frac{1+\epsilon}{\sqrt{\epsilon^2 + 2(1+\epsilon)(1-cos\theta)}}, \tan\phi = -\frac{sin\theta}{1 + \epsilon - cos\theta}, \quad (6)$$

Let $\boldsymbol{g}_i \cdot \boldsymbol{R}_j = 2\pi\delta_{ij}$ be the corresponding moiré reciprocal lattice. Geometrically, we consider hBN on layer $l = 0$ underneath layer $l = 1$ of R5LG. There are two options for the sublattice alignment, labelled by $\xi$. When $\xi = 0$, then $(B, N)$ are aligned directly underneath the $(A_1, B_1)$ carbon atoms in the AA region of the moiré unit cell; for $\xi = 1$, $(N, B)$ are underneath $(A_1, B_1)$ in AA.

For the moiré continuum model [9,32] we take a basis $|\boldsymbol{k}, \sigma, l, \tau, s\rangle$ where $l \in \{0,1,...,5\}$, $\tau = \{K, K'\}$ is valley, $s = \{\uparrow, \downarrow\}$ is spin, and $\boldsymbol{k}$ is in the moiré Brillouin zone. In the $(K, \uparrow)$ sector, the continuum model is

$$h^{(\xi)}_{K\uparrow} = \begin{bmatrix} \sigma_x^\xi \hat{h}^{hBN} \sigma_x^\xi & \hat{T}^{(\xi)} \\ (\hat{T}^{(\xi)})^\dagger & \hat{h}^{R5G}(\boldsymbol{k}) \end{bmatrix}, \quad (7)$$

Where $\xi = 0,1$ and $\sigma^x$ is the Pauli matrix. The tunneling is entirely specified by the 2x2 block between $l = 0$ and 1

$$T^{(\xi)}(\boldsymbol{r}) = T_0^{(\xi)} + T_1^{(\xi)} e^{i\boldsymbol{g}_1 \cdot \boldsymbol{r}} + T_2^{(\xi)} e^{i(\boldsymbol{g}_1 + \boldsymbol{g}_2) \cdot \boldsymbol{r}}, \quad (8)$$

where

$$T_n^{(\xi)} = \sigma_x^\xi \begin{bmatrix} t_{BC} & 0 \\ 0 & t_{NC} \end{bmatrix} \sigma_x^\xi \begin{bmatrix} \kappa_{hBN} & \omega^n \\ \omega^{-n} & \kappa_{hBN} \end{bmatrix}, \quad (9)$$

with $\omega = e^{-i2\pi/3}$ is the third root of unity. Following [32], we use $r_{NB} = t_{NC}/t_{BC}$, the ratio of nitrogen-carbon tunneling to boron-carbon tunneling, as well as $\kappa_{hBN} \in [0,1]$. The latter reduces tunneling in the AA region, phenomenologically accounting for the shrinking of the AA regions under lattice relaxation. As usual, the K' valley is defined by the action of time-reversal and spin ↓ is determined by $SU(2)$ spin symmetry. All parameters are given in Extended Data Table 2.

| $t_0$ | $t_1$ | $t_2$ | $t_3$ | $t_4$ | $\delta_1$ | $\delta_2$ | $t_{BC}$ | $r_{NB}$ | $\kappa_{hBN}$ |
|---|---|---|---|---|---|---|---|---|---|
| 3100 | 380 | -15 | -290 | -141 | 10.5 | 2 | 144 | 0.5 | 0.5 |

**Extended Data Table 2.** Parameters for the R5LG/hBN moiré model. All entries are given in meV, except for the dimensionless $r_{NB}$ and $\kappa_{hBN}$.

**Stoner model**

The large density of states at vHs of the conduction band under displacement field give rise to Stoner ferromagnetism. To capture this, we follow an approach that has been successfully applied to R3LG [3], as well as to twisted bilayer graphene [63] and twisted trilayer supermoiré systems [64].

Our starting point is the interacting continuum model in the band basis:

$$\hat{H} = \sum_{k,b,\alpha} (\epsilon_{k\alpha} - \mu)\hat{c}^\dagger_{kb\alpha}\hat{c}_{kb\alpha} + \frac{1}{2}\int drdr'\, V(r-r') : \hat{\rho}(r)\hat{\rho}(r') : \quad (10)$$

where $\hat{c}^\dagger_{kb\alpha}$ creates a fermion in band $b$ with flavor $\alpha$ at point $k$ in the mBZ, $\epsilon_{kb\alpha}$ is the corresponding single-particle energy, $\mu$ is a chemical potential, $\hat{\rho}(r)$ is the density at $r$ from all flavors, $V$ represents density-density interactions, and normal ordering :: is with respect to charge neutrality.

Following [63], we consider a Stoner model with three simplifying assumptions: $(A1)$ contact interactions with $(A2)$ trivial interaction matrix elements and $(A3)$ meanfield theory. From $(A1)$, interactions must be strictly flavor-off-diagonal (so the Fock term exactly vanishes). As the form factors are trivial by $(A2)$, and the interaction matrix elements are $q$-independent due to contact interactions, the $k$-dependence is irrelevant. The problem is therefore totally specified by the density of states of each flavor, and Slater determinants $|\Psi_{MF}(\{n_\alpha\})\rangle$ are specified by the density $n_\alpha$ in each flavor $\alpha$. Under interactions, the bands of this model may shift up and down rigidly, producing ferromagnetism. Within-band deformations - ubiquitous in such strongly interacting systems - are forbidden here.

Under these assumptions, the grand potential per unit area is

$$\frac{\Phi_{MF}}{A} = \sum_\alpha E(n_\alpha) + \frac{A_{uc}}{2}\sum_{\alpha \neq \beta} U_{\alpha\beta} n_\alpha n_\beta, \quad (11)$$

where the density and kinetic energy

$$n_\alpha = \int_0^{\mu(n_\alpha)} \rho_\alpha(E)dE, \quad E(n_\alpha) = \int_0^{\mu(n_\alpha)} \varepsilon\rho_\alpha(E)dE \quad (12)$$

are determined from the density of states

$$\rho_\alpha(E) = \frac{1}{N}\sum_{k,b} \delta(\varepsilon - \varepsilon_{kb\alpha}). \quad (13)$$

Here $A_{uc} = \frac{\sqrt{3}}{2}a^2$ is the graphene unit cell area, $N$ is the number of unit cells and $A = NA_{uc}$. Due to $SU(2)$ spin and $U(1)$ valley symmetry, the density of states is identical in each flavor, whereupon flavor-isotropic interactions $U_{\alpha\beta} = U(1 - \delta_{\alpha\beta})$ would give $SU(4)$ symmetry. Experiments (and more sophisticated theory) indicate that increasing density first breaks spin symmetry then valley symmetry. To reproduce this phenomenology we add a Hund's term $J$ [3] to get an interaction term

$$\frac{A_{uc}}{2} \sum_{\alpha \neq \beta} U n_\alpha n_\beta - J A_{uc}(n_1 - n_3)(n_2 - n_4), \quad (14)$$

where $\alpha, \beta = \{1,2,3,4\} = \{K\uparrow, K'\uparrow, K\downarrow, K'\downarrow\}$. At each $\mu$, the ground state is easily found from the grand potential minima at

$$0 = \frac{d}{dn_\alpha}\frac{\Phi_{MF}}{A} = \mu_\alpha(n_\alpha) + A_{uc}\sum_{\beta \neq \alpha} U_{\alpha\beta} n_\beta - \mu, \quad (15)$$

where $\mu_\alpha(n_\alpha) = \frac{dE(n_\alpha)}{dn_\alpha}$. We take identical parameters to those used in non-moiré R3LG: $U = 15\ eV, J = 4.5\ eV$ [3]. This leads to the Hund's coupling $J_H = J \cdot A_{uc} = 2.4 \times 10^{-12}\ meV \cdot cm^{-2}$, comparable to the value derived in [5].

Numerically, we restrict to the lowest seven conduction bands and use $64 \times 64\ k$-points per valley to resolve vHs. We report the density in units of moiré filling, e.g. $\nu = 1$ corresponds to $n \approx 0.64 \times 10^{-12}\ cm^{-2}$ at $\theta = 0.3°$. To reliably find the global minima, we use at least 10 initial states for each minimization.

**Calculation of orbital magnetization**

We compute the orbital magnetization $M(n)$ as a function of carrier density using an approximate method, since no practical formula exists for interacting systems. Our approach combines the single-particle orbital magnetization calculation with flavor-resolved densities obtained from the Stoner model. After calculating the occupation $n_\alpha$ of each flavor, we approximate the total magnetization $M(n)$ at a given total density $n = \sum_\alpha n_\alpha$ by

$$M(n) = \sum_\alpha \tau_\alpha \widetilde{M}(n_\alpha(\mu_\alpha)) + s_\alpha \mu_B, \quad (16)$$

where $\tau_\alpha = \pm 1$ is valley, $s_\alpha = \pm 1$ is spin, $\mu_B$ is the Bohr magneton, and $\widetilde{M}(\mu_\alpha) = M_{SP}(\mu_\alpha) - M_{ref}(\mu_\alpha)$. Here the single-particle orbital magnetization is

$$M_{SP}(\mu_\alpha) = \frac{e}{\hbar}\varepsilon^{ijz}\sum_b \int \frac{d^2\mathbf{k}}{(2\pi)^2} f(\varepsilon_{\mathbf{k}b} - \mu_\alpha) \times \langle \partial_{k_i} u_{\mathbf{k}b\alpha} | h_{\mathbf{k}\alpha} + \varepsilon_{\mathbf{k}b\alpha} - 2\mu_\alpha | \partial_{k_j} u_{\mathbf{k}b\alpha} \rangle, \quad (17)$$

where $f$ is the $T = 0$ Fermi factor, $u_{\mathbf{k}b\alpha}$ is the periodic part of the Bloch wavefunction, $h$ is the single-particle Hamiltonian for flavor $\alpha$, and $i,j \in \{x,y\}$ are spatial indices with an implicit sum. Finally, $M_{ref}(\mu_\alpha)$ is chosen to be an $\alpha$-independent linear function of $\mu_\alpha$ so that $M(n)$ is zero within the gap at charge neutrality. The chosen "reference subtraction" corresponds to the orbital magnetization generated by moving the valence bands of the other flavors down by $\mu_\alpha$. Given the magnetization $M(n)$, we compute $dM/dn$, which is show in Figs. 3c,f.

## Single particle hBN-moiré parameters

To highlight the importance of using accurate hBN moiré parameters—specifically $r_{NB}$ and $\kappa_{hBN}$—we compare calculated DOS maps for both alignment orientations $\xi$, using all four combinations of $r_{NB} = 0.5, 1$ and $\kappa_{hBN} = 0.5, 1$. Extended Data Fig. 6 presents the results for each parameter set and each $\xi$. When both parameters are set to 1 (panels d,h), effectively neglecting moiré relaxation and assuming equal tunneling amplitudes between carbon–boron and carbon–nitrogen, the two alignment orientations yield nearly identical DOS maps, with only subtle and difficult-to-resolve differences. Introducing relaxation alone ($\kappa_{hBN} = 0.5$, $r_{NB} = 1$, panels c,g) leads to pronounced differences between $\xi = 0$ and $\xi = 1$; however, this configuration predicts a persistent gap at $\nu = -4$ across a wide range of moiré-proximate displacement fields in particular for $\xi = 0$, inconsistent with experimental observations. Conversely, enabling only the tunneling asymmetry by setting $r_{NB} = 0.5$ while keeping $\kappa_{hBN} = 1$ (panels b,f) again results in nearly indistinguishable DOS maps for the two alignment orientations, and fails to capture the observed asymmetry. Only when both effects are included—using the experimentally derived parameter set [32] with $r_{NB} = 0.5$ and $\kappa_{hBN} = 0.5$ (panels a,e)—do we reproduce both key features: a clear distinction between the two $\xi$ configurations and the absence of a robust gap at $\nu = -4$ for $\xi = 0$ on the hole-doped, moiré-proximate side.

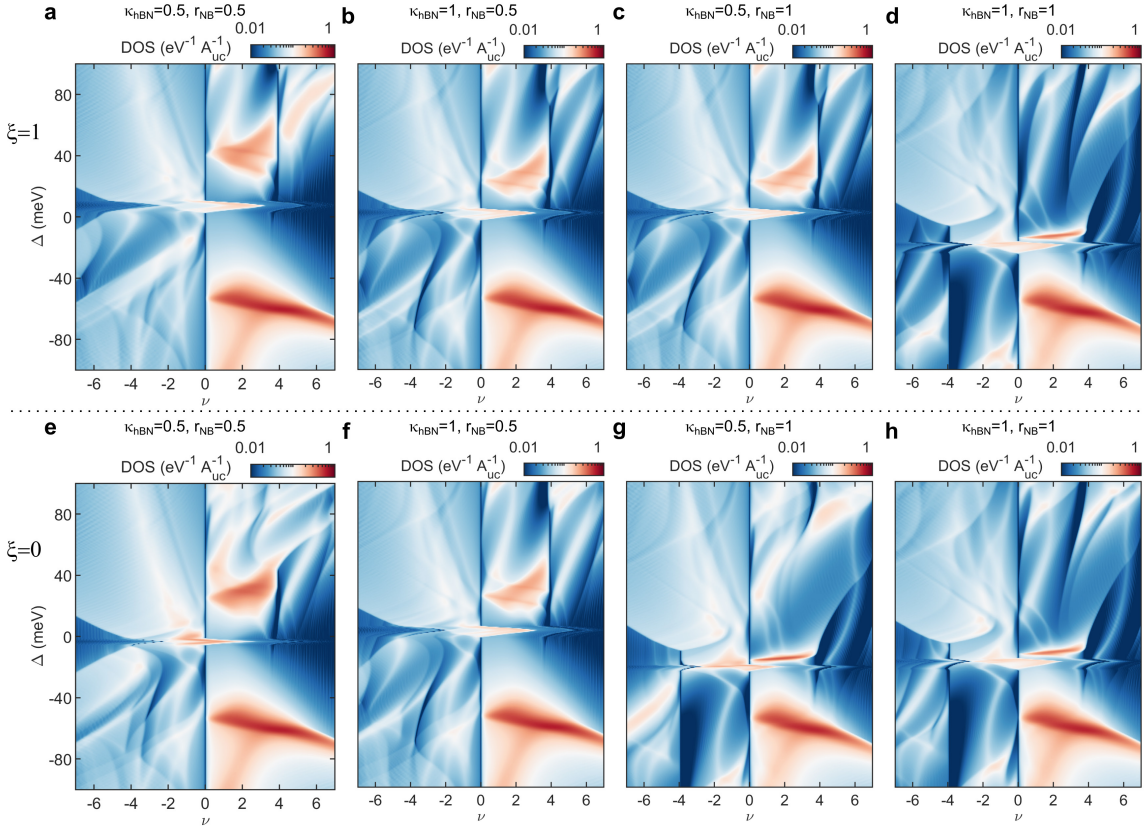

**Extended Data Fig. 6. Dependence of DOS maps on $\kappa_{hBN}$ and $r_{NB}$. a-d**, Calculated DOS maps for $\xi = 1$ using different combinations of $\kappa_{hBN}$ and $r_{NB}$: (**a**) $\kappa_{hBN} = 0.5$, $r_{NB} = 0.5$; (**b**) $\kappa_{hBN} = 1$, $r_{NB} = 0.5$; (**c**) $\kappa_{hBN} = 0.5$, $r_{NB} = 1$; (**d**) $\kappa_{hBN} = 1$, $r_{NB} = 1$. **e-h**, Same as (a-d) for $\xi = 0$. Including both moiré relaxation ($\kappa_{hBN} < 1$) and tunneling asymmetry ($r_{NB} < 1$) is essential to capture the observed differences in alignment-dependent DOS features.

## Weakly-magnetic phase

The SOT magnetometry data of the $\xi = 1$ device in Fig. 3a shows a clear phase transition from a valley- and spin-polarized (VP-SP) phase to a different state with only weak magnetization, occurring around $D \approx 0.4$ V/nm. This WM phase exhibits a pronounced resistance peak at $\nu = 1$, and has therefore been suggested to correspond to an IVC state. While this interpretation appears to be the most likely, the experimental evidence remains insufficient to conclusively verify it, and alternative explanations cannot be ruled out.

One such alternative is that the region marked in light blue in Fig. 3b does not represent a single uniform phase, but rather a broader domain in which the relative polarization between spin and valley evolves continuously as a function of $\nu$ and $D$. In this scenario, a gap opens only near $\nu = 1$ where the system reaches a VP-SP configuration. The absence of a sharp $dM/dn$ peak across this transition could then be attributed to a gradual crossover, with magnetization changes falling below our experimental resolution.